\documentclass[energies,article,accept,moreauthors,pdftex]{Definitions/mdpi}

\firstpage{1}
\pubvolume{13}
\issuenum{4}
\articlenumber{988}
\pubyear{2020}
\copyrightyear{2020}
\history{Received: 30 January 2020; Accepted: 19 February 2020; Published: 22 February 2020}
\updates{yes} 

\usepackage{acronym}
\acrodef{aoe}[AOE]{average objective error}
\acrodef{atr}[ATR]{average time reduction}
\acrodef{dlrve}[DLR-VE]{German Aerospace Center---Institute of Networked Energy Systems}
\acrodef{etrago}[eTraGo]{Electric Transmission Grid Optimization}
\acrodef{lopf}[LOPF]{linear optimal power flow}
\acrodef{nep}[NEP 2035]{Netzentwicklungsplan 2035}
\acrodef{milp}[MILP]{mixed integer linear program}
\acrodef{openego}[open\_eGo]{Open Electricity Grid Optimization project}
\acrodef{oep}[OEP]{Open Energy Platform}
\acrodef{pypsa}[PyPSA]{Python for Power System Analysis}
\acrodef{rmse}[RMSE]{root mean square error}
\acrodef{tsam}[TSAM]{time series aggregation method}

\usepackage{xcolor}

\usepackage{footnote}

\usepackage[binary-units=true,group-separator={,}]{siunitx}
\DeclareSIUnit\EUR{EUR}
\DeclareSIUnit\watthour{Wh}
\DeclareSIUnit\voltampere{VA}
\DeclareSIUnit{\million}{\text{million}}

\graphicspath{{Pictures/}}

\Title{Evaluation of Temporal Complexity Reduction Techniques Applied to Storage Expansion Planning in Power System Models}

\Author{{Oriol Ravent}{\'o}s {*}\href{https://orcid.org/0000-0002-0512-4331}{\orcidicon} and Julian Bartels\href{https://orcid.org/0000-0002-2623-9787}{\orcidicon}}

\AuthorNames{Oriol Ravent{\'o}s and Julian Bartels}

\address[1]{%
DLR--Institut f{\"u}r Vernetzte Energiesysteme e.V., Carl-von-Ossietzky-Str.\ 15, 26129 Oldenburg, Germany; julian.bartels@dlr.de 
}

\corres{\hangafter=1 \hangindent=1em \hspace{-1em} Correspondence: oriol.raventosmorera@dlr.de 
}

\abstract{The growing share of renewable energy makes the optimization of power flows in power system models computationally more complicated, due to the widely distributed weather-dependent electricity generation.~This article evaluates two methods to reduce the temporal complexity of a power transmission grid model with storage expansion planning.~The goal of the reduction techniques is to accelerate the computation of the linear optimal power flow of the grid model. This is achieved by choosing a small number of representative time periods to represent one whole year. To select representative time periods, a hierarchical clustering is used to aggregate either adjacent hours chronologically or independently distributed coupling days into clusters of time series.~The~aggregation efficiency is evaluated by means of the error of the objective value and the computational time reduction. Further, both the influence of the network size and the efficiency of parallel computation in the optimization process are analysed. As a test case, the~transmission grid of the northernmost German federal state of Schleswig-Holstein with a scenario corresponding to the year 2035 is considered.~The considered scenario is characterized by a high share of installed~renewables.}


\begin{document}


\section{Introduction}

One of the most accepted key strategies to address Global Warming is the substantial increase of the renewable energy capacity. This requires a significant transformation of the power network, from few stable big power plants to many weather-fluctuating widely\nobreakdash-distributed small power sources. Therefore, in order to achieve a successful transition of the energy system, a detailed planning of the power grid to distribute energy as well as storage planning to ensure a constant power supply are required. The~main objective of this article is to evaluate different {\emph{time series aggregation methods}} (TSAMs) to reduce the computational time of the \acfi{lopf}, which is the linear approximation of the power flow equations while minimizing the total system costs. A \acs{tsam} approximates the input time series for one whole year, i.e., the~load and the renewable energy generation, by a small set of representative time periods (in our case hours or days). The~procedure used in this paper involves the following steps: scale and normalize the time series, cluster the time series in groups of periods, choose a representative of each cluster, adapt the corresponding equations on the linear power flow problem, and rescale the representative time series. The~necessary tracking of the state of charge of the storage units makes the implementation of the time series clustering more challenging, since it adds constraints linking different time periods. Moreover, the~high share of unpredictable renewable energy sources leads to a potentially less accurate approximation of the renewable energy time series. 

The paper is structured as follows. In Section \ref{literature} we conduct a brief literary review of temporal complexity reduction of power grids focusing on the methods that use representative periods. In~the methodology Section \ref{sec_methodology}, the~test case, the~linearization of the power flow, the~clustering methods, the~indicators, and the software and hardware used are described in detail. Section \ref{sec_test_opti} contains the optimization of the reference test case. Sections \ref{sec_tsam_efficiency} and \ref{sec_time_red_lopf} contain the main results of the paper, comparing the different \acsp{tsam}. Sections \ref{sec_size} and \ref{sec_CPU} study the influence of the network size and the parallel computation on the results. Finally, the~conclusions are drawn in Section \ref{sec_conclusion}.

\section{Literature Review} \label{literature}
In this section, we conduct a brief literary review of temporal complexity reduction of power grids focusing on the methods that use representative periods. In \cite{Po17} some methods are presented to select representative periods and some indicators are introduced to evaluate the accuracy of the time series representations. Those methods include heuristics, clustering algorithms, random selection{,} \acfi{milp}{,} and hybrid approaches. However, the~methods are not evaluated in terms of the accuracy or the time reduction of the linear optimization of the network. Nevertheless, a~\acs{milp} to optimize the temporally aggregated network is stated, but, as noted in \cite{Na16} (page~431), for~large networks a \acs{milp} would be computationally too complicated to solve. Most of the recent articles essentially follow the clustering method presented in \cite{Na16}, specially regarding the normalization and the rescaling of the data. They use the hierarchical clustering with decoupled periods, which~amounts to consider each cluster of time series as an isolated entity and, afterwards, paste the results of all clusters together. This method is prone to highly underestimate the storage capacity, since it does not have a coherent state of charge for interperiod storage. For this reason, in \cite{Ga18}, the~coupling periods method is introduced, which adds extra equations to the decoupling periods method to keep track of the storage units state of charge. In~\cite{Ko18b}, different clustering algorithms are evaluated, like the $k$-means clustering, the~hierarchical clustering, and the hierarchical clustering with added load and feed-in peak periods. In a succeeding article \cite{Ko18}, two different sets of equations are introduced to define the linear problem, in order to simplify the coupling periods method for days by the introduction of new variables to take care of the intra\nobreakdash-day state of charge and the inter\nobreakdash-day state of charge separately. This results in a remarkable computational time reduction for an island energy system (but with no storage planning). A similar method, clustering periods of hours and adding a different set of equations to keep track of the state of charge, is presented in \cite{Te18} as a system states model with a reduced frequency matrix. Finally, in \cite{Pi18}, the~time series aggregation method of chronological periods is used, which forces the clustering algorithm to cluster just adjacent time periods. This approach remarkably simplifies the equations taking care of the state of charge, hence allowing a much faster computation.

In \cite{Pri19}, a method is presented to systematically analyze the trade\nobreakdash-off between complexity and accuracy of energy system models. This is useful in order to anticipate if a certain complexity reduction technique would have a big influence in the computational time and the deviation from the benchmark. Regarding temporal resolution, they conclude that it is strongly related to the solving time. It is also observed \cite{Pri19} (page~21), that the temporal resolution has a stronger impact on the accuracy of the optimization than the spatial resolution, specially if the model have temporal linking constraints, like~ramping or dynamic state of charge (as in our case).

The present contribution compares the clustering of coupling days non-chronologically \cite{Ga18}, which~will be denoted here by {coupling clustering}, with the clustering of the hours chronologically \cite{Pi18}, denoted here by {chronological clustering} in the context of storage expansion planning. The~comparison will be in terms of computational time and error. In \cite{Pi18}, the~chronological clustering is not compared to the coupling clustering, but it is shown to be more efficient than the decoupling periods method~\cite{Na16}, which ignores the consistency of the state of charge. This result is expectable in a system with a large storage capacity.~To our knowledge, the~present is the first article which compares the chronological clustering to another method that keeps a consistent state of charge.~Other clustering methods (like~$k$-means), other linear problem implementations (with different constraint formulations), or other types of periods (like~weeks) will not be considered here, since we set the scope of this study to the two methods known to be computationally the more robust and efficient, \mbox{{cf.}\ \cite{Na16,Ga18,Ko18b,Ko18,Pi18,Pf17}}. The~approaches used will be evaluated on a reference energy system developed in the \acfi{openego} \cite{openego}. The~linear optimization problem is solved by the open source Python package \acfi{pypsa} \cite{pypsa}. In~contrast to many of the publications mentioned above, which consider between 10 and 50 time series, our test case uses 437~time series generating a linear problem with up to 6 million variables and 6 million equations. In~addition, this contribution compares the computational time reduction (not only the precision of the approximation) and, further, analyses the influence of the size of the network and the potential of parallel computation.

\section{Methodology}\label{sec_methodology}

In this section we present our methodology, which consists of a test case, the~\acs{lopf}, the~clustering methods that will be evaluated, the~indicators, and the hardware and software used in our experiments.

\subsection{Test case}\label{sec_tests}

\begin{figure}[H]
\centering
\includegraphics[width=0.95\linewidth]{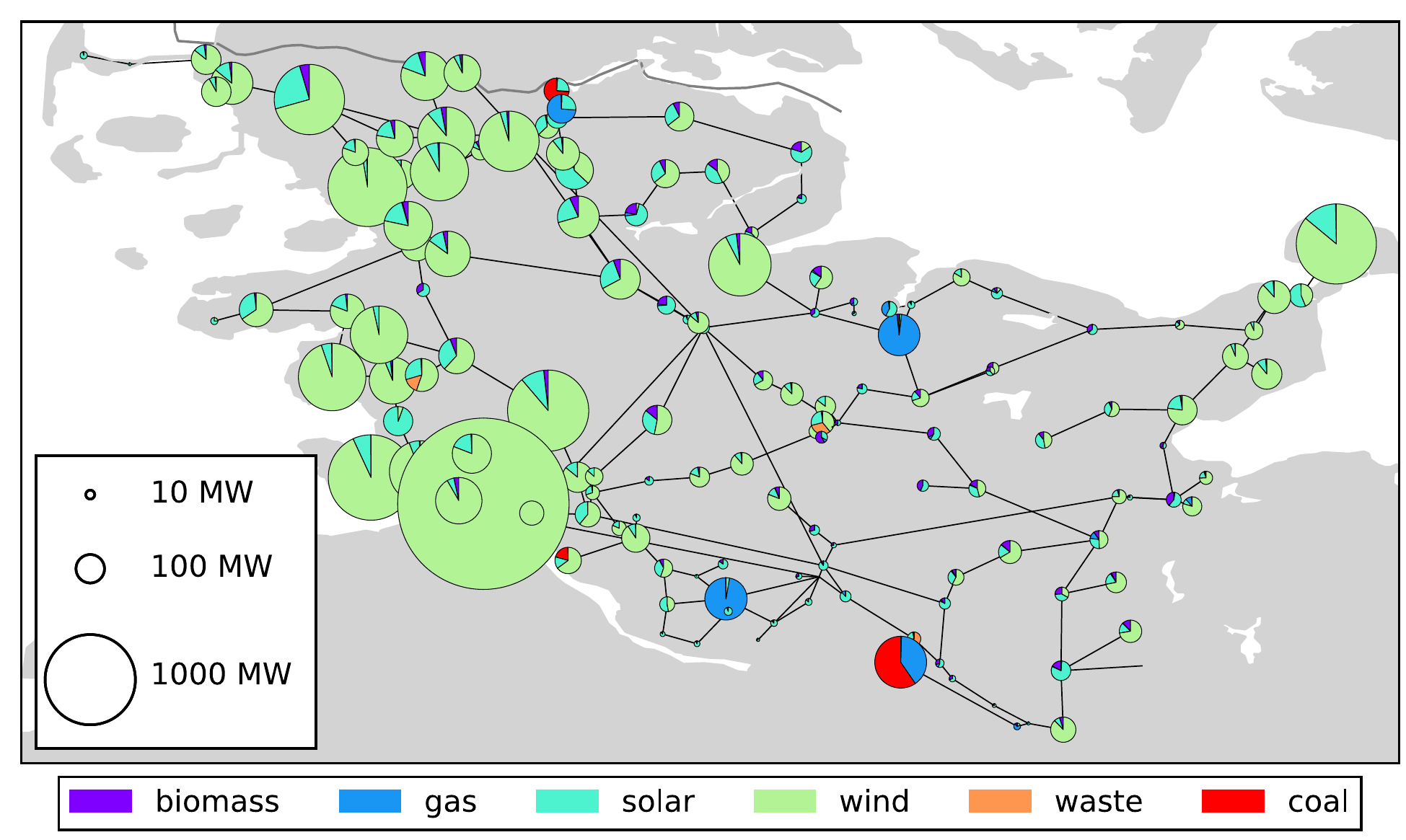}
\caption{Reference power grid in Schleswig-Holstein with installed capacities.}\label{fig_grid}
\end{figure}

As a test case, we use a 152 node version of the German federal state of Schleswig-Holstein power transmission grid in an exogenous scenario for 2035 as described in \cite{osf}. This data is publicly available and we use it via the open source Python package \acs{etrago} \cite{etragodoc} (using the grid version ``0.2.11'' and the scenario ``SH NEP 2035''). This power system is characterized by the high availability of wind energy and the possibility of expanding the storage (both interday and intraday) almost infinitely at a competitive annualized cost, i.e.,\ it is a {\emph{storage capacity expansion planning problem}}. Both~the high wind availability and storage expansion option, make the system computationally more challenging to optimize.  The spatial clustering of the network to 152 nodes is done using the \acs{pypsa} tool ``kmeans\_clustering''. A plot of the resulting network is shown in Figure \ref{fig_grid}, where the size of each bus is proportional to its installed capacity. Table \ref{table_capacity} contains the global installed capacity used in the~model.

The main characteristics of the power grid and its components are summarized in Table \ref{table_grid} after the references in \cite{osf,TSO15,El16}. Notice that, in the model, no capital costs are considered for generators, i.e.,~its installed capacity is fixed. Note also that wind and solar energy generation are assumed to have no operational costs. The~properties that will most certainly influence the \acs{lopf} are the large installed capacities of wind energy (due to its unpredictable nature) and the option to extend the storage units capacity almost endlessly at a competitive (annualized) price. There are two types of storage systems considered in the model: batteries with a round-trip efficiency of $\eta^{dis} \eta^{char}=$ {87}\% and a smaller power input, functioning as intra\nobreakdash-day storage (with an energy to power ration of $q=6$) and compressed hydrogen stored in underground salt caverns functioning as inter\nobreakdash-day storage (with an energy to power ration of $q=168$). This latter has a very low round-trip efficiency of {30}\%, but has a much bigger power input.

\unskip
\begin{table}[H]
\centering
\caption{{Installed capacity by carriers} (\SI{}{\giga\watt}) provided by the data of \cite{openego} from  \cite{opsd}.}



\label{table_capacity}
\begin{tabular}{cccccc}
\toprule
\bf{Wind} & \bf{Solar} & \bf{Gas} & \bf{Waste} & \bf{Coal} & \bf{Biomass} \\
\midrule
\num{15.56} & \num{2.57} & \num{0.68} & \num{0.05} & \num{0.27} & \num{0.5} \\
\num{79.17}\% & \num{13.13}\% & \num{3.49}\% & $<$\num{0.01}\% & \num{1.36}\% & \num{2.56}\% \\
\bottomrule
\end{tabular}
\end{table}
\unskip

\begin{table}[H]
\caption{{Main characteristics of} the power grid components. Generator costs ({including fuel and} $\text{CO}_2$ prices as well as variable and fixed costs) are derived by \cite{osf}, storage units parameter by \cite{TSO15,El16}.}
\label{table_grid}
\centering
\begin{tabular}{lc l}
\toprule
\bf{Component} & \bf{Quantity} & \multicolumn{1}{c}{\bf{Characteristics}} \\
\midrule
\bf{Buses} & 152 & \SI{380}{\kilo\volt} (unified by the clustering) \\\midrule
\bf{Lines} & 132 & between \num{182} and \SI{3234}{\mega\voltampere} \\\midrule
\bf{Snapshots} & 8760 & \\\midrule
\bf{Loads} & 149 & \\\midrule
\bf{Generators} &  & fixed capacity  \\
\;\; Biomass & 125 & \SI{31.4112}{\EUR\per\mega\watthour} \\
\;\; Gas & 125  &  \SI{41.9344}{\EUR\per\mega\watthour} \\
\;\; Wind & 115 & \SI{0}{\EUR\per\mega\watthour} \\
\;\; Solar & 146 & \SI{0}{\EUR\per\mega\watthour} \\
\;\; Waste & 3 & \SI{39.463}{\EUR\per\mega\watthour} \\
\;\; Coal & 3 & \SI{24.7914}{\EUR\per\mega\watthour} \\ \midrule

\bf{Storage units} &  & {extendable} up to \SI{1}{\tera\watt}	\\

\;\; Batteries  & 149 & \SI{65822}{\EUR\per\mega\watt}, \SI{0.01}{\EUR\per\mega\watthour},\\ & & $\eta^{loss}=$ \num{0.00694}, $\eta^{dis}=\eta^{char}=$ \num{0.9327}, $q=6$ \\
\;\; Hydrogen & 51 & \SI{65402}{\EUR\per\mega\watt}, \SI{0.01}{\EUR\per\mega\watthour},\\
&  & $\eta^{loss}=$ \num{0.000694}, $\eta^{dis}=$   \num{0.425}, $\eta^{char}=\num{0.725}$, $q=168$ \\
\bottomrule
\end{tabular}
\end{table}

\subsection{Linear Optimization Power Flow}\label{sec_lopf}
The linear problem computing the \acs{lopf} is implemented using the free software toolbox \acs{pypsa}~\cite{pypsa}.~It is well-known that the linearization of the power flow equations introduces only negligible errors in the case that the voltage angle differences across the branches are small, the~branch resistances are negligible compared to their reactances, and voltage magnitudes can be kept at nominal values (which is the case when simulating transmission systems). From the many equivalent formulations to compute the \acs{lopf} in \acs{pypsa}, the~{\emph{Angles+Flow multi-period formulation with long term storage}} will be used here. Next, we list all the equations of this formulation using the variables and parameters as defined in Table \ref{table_variables}. For a detailed formulation we refer to \cite{Ho18}.

\begin{table}[H]
\caption{Variables, coefficients and indices of the Angles+Flow multi-period with long term storage formulation of the \acs{lopf}.}
\label{table_variables}
\centering
\begin{tabular}{ l l}
\toprule
& \multicolumn{1}{c}{\bf{Definition}} \\
\midrule
$i\in\{1,\dots, I\}$ &  bus label \\
$\ell\in\{1,\dots, L\}$ & line label \\
$s\in\{1,\dots, S\}$ & generator label \\
$r\in\{1,\dots, R\}$ & storage unit label \\
$t\in\{1,\dots, T\}$ & snapshot or time step \\
$\omega_t$ & weighting of a snapshot (\si{\hour}) \\
$g_{i,s,t}$   & generator dispatch (\si{\mega\watt}) \\
$G_{i,s}$ &  generator power capacity (\si{\mega\watt}) \\
$\tilde{G}_{i,s}$, $\bar{G}_{i,s}$ &  minimum and maximum installable generator potential (\si{\mega\watt}) \\
$\tilde{g}_{i,s,t}$, $\bar{g}_{i,s,t}$ &  minimum and maximum generator power availability $\in(0,1)$ \\
$o_{i,s}$ & operating cost of a generator (\si{\EUR\per\mega\watthour}) \\
$c_{i,s}$ &  capital cost of a generator (\si{\EUR\per\mega\watt}) \\
$\theta_{i,t}$ &  voltage angle at a bus (\si{\radian}) \\
$f_{\ell,t}$ &   power flow at a line (\si{\mega\watt}) \\
$F_{\ell}$   & power rating at a line (\si{\mega\watt}) \\
$K$ &  $I\times L$ incidence matrix \\
$B$ &  diagonal $L\times L$ matrix of line susceptances \\
$p_{i,t}$ &  total active power injection at a bus (\si{\mega\watt}) \\
$l_{i,t}$ &  load (\si{\mega\watt}) \\
$h_{i,r,t}$ &  dispatch of a storage unit (\si{\mega\watt}) \\
$H_{i,r}$ &  power capacity of a storage unit (\si{\mega\watt}) \\
$\tilde{H}_{i,r}$, $\bar{H}_{i,r}$ &  installable potential of a storage unit (\si{\mega\watt}) \\
$\tilde{h}_{i,r,t}$, $\bar{h}_{i,r,t}$ &   power availability per unit of storage capacity \\
$o_{i,r}$ &   operating costs of a storage unit (\si{\EUR\per\mega\watthour}) \\
$c_{i,r}$ &  capital cost of a storage unit (\si{\EUR\per\mega\watt}) \\
$e_{i,r,t}$ &  state of charge of a storage unit (\si{\mega\watthour}) \\
$q_{i,r}$ &  hours at nominal power to fill up a storage unit, i.e.,\ energy to power ratio, (h)\\ 
$\eta^{loss}_{i,r}$ & storage unit losses per hour\\
$\eta^{char}_{i,r}$ & efficiency of charge of a storage unit \\
$\eta^{dis}_{i,r}$ & efficiency of discharge of a storage unit \\
\bottomrule
\end{tabular}
\end{table}

The objective function minimizing the total costs is:
\begin{equation}\label{eq-gen-obj}
\mbox{\rm min} \left[ \sum_{t} \left(\omega_t \left(\sum_{i,s} o_{i,s} g_{i,s,t} + \sum_{i,r} o_{i,r} \left[h_{i,r,t}\right]^+  \right) \right) +  \sum_{i,s} c_{i,s} G_{i,s} + \sum_{i,r} c_{i,r} H_{i,r} \right],
\end{equation}
where, the~weighting is such that $\sum_{t} \omega_t = 8760$. The~buses power balance is defined by:
\begin{equation}\label{c1}
p_{i,t}=\sum_{s} g_{i,s,t} + \sum_{r} h_{i,r,t} - l_{i,t} \; \forall i \mbox{ and } t,
\end{equation}
with power dispatch bounds:
\begin{equation}\label{c2}
\tilde{g}_{i,s,t}G_{i,s}\leq g_{i,s,t}\leq \bar{g}_{i,s,t} G_{i,s} \; \forall i, s \mbox{ and } t.
\end{equation}

The power flow limits are expressed by:
\begin{equation}\label{c22}
|f_{\ell,t}|\leq F_\ell \; \forall \ell \mbox{ and } t
\end{equation}
and the Kirchhoff current laws take the form:
\begin{equation}\label{c3}
f_{\ell,t}=\sum_{i} \left(B K^{T}\right)_{\ell i}\theta_{i,t} \; \forall \ell \; \mbox{\rm and}\; t,
\end{equation}
\begin{equation}\label{c4}
p_{i,t} = \sum_{\ell} K_{i \ell} f_{\ell,t} \; \forall i \;\mbox{\rm and}\; t \mbox{\rm , and}
\end{equation}
\begin{equation}\label{c42}
\theta_{0,t}=0 \; \forall t.
\end{equation}

We also consider the following constraints:
\begin{equation}\label{c5}
\tilde{G}_{i,s}\leq G_{i,s} \leq\bar{G}_{i,s} \;\forall\; i, s,
\end{equation}
\begin{equation}\label{c6}
\tilde{h}_{i,r,t}H_{i,r}\leq h_{i,r,t}\leq \bar{h}_{i,r,t} H_{i,r} \; \forall r, i \mbox{ and } t,
\end{equation}
\begin{equation}\label{c7}
\tilde{H}_{i,r}\leq H_{i,r} \leq\bar{H}_{i,r} \;\forall\; i, r,
\end{equation}
\begin{equation}\label{c8}
0\leq e_{i,r,t}\leq q_{i,r}H_{i,r} \; \forall i, r \mbox{ and } t,
\end{equation}
and the \emph{time linking constraint} keeping track of the state of charge:
\begin{equation}\label{c9}
e_{i,r,t} = (1-\eta^{loss}_{i,r})^{\omega_t} e_{i,r,t-1} + \omega_t\left(\eta^{char}_{i,r} \left[h_{i,r,t}\right]^+ - \frac{1}{\eta^{dis}_{i,r}}\left[h_{i,r,t}\right]^-\right) \; \forall i, r \mbox{ and } t,
\end{equation}
assuming a cyclic indexing, i.e.,\ $e_{i,r,0} = e_{i,r,T}$. That means that the state of charge is forced to be the same at the beginning and at the end of the year. Notice that the weighting $\omega_t$ as an exponent of the losses means that we account for the accumulated losses for each hour in $\omega_t$ with respect to the previous state of charge (but assuming no losses of the charged and discharged energy in the those internal hours).

Notice that the only time linking constraint, i.e.,\ containing variables depending on both $t$ and $t-1$, is Equation (\ref{c9}). This means that only this equation has to be modified after the time series aggregation, in order to keep the time coherence.

\newpage
\subsection{Time Series Aggregation Techniques}\label{sec_tsam}

Next we describe in details the two \acsp{tsam} evaluated in this paper. The~open source packages tsam~\cite{tsam}, and eTraGo~\cite{etragodoc}, were used to perform the time series aggregations and to adapt the test case network. The~packages were conveniently adapted to perform the chronological clustering, the~linear problem formalization from \cite{Ko18}, and a more elaborate data pre-processing.

\subsubsection{Data Preprocessing}\label{sec_pretreat}

The first step of a \acs{tsam} is to normalize the different time series so that the clustering would not be affected by the different orders of magnitudes of the time series. The~load is normalized with respect to the maximum load value at each grid bus, whereas the energy generation is normalized across all buses, cf.\ \cite{Na16}. Then, the~time series for the load, wind and solar energy generators at all buses are concatenated in a single time series $\{x_t\}_{t = 1,\dots,8760}$ with $x_t = (l_{1,t},\dots,l_{I,t},\tilde{g}_{1,wind,t},\dots,\tilde{g}_{I,wind,t},\tilde{g}_{1,solar,t},\dots,\tilde{g}_{I,solar,t})$.

\subsubsection{Clustering}\label{sec_clustring}

The \emph{coupling clustering} for days groups the time series of the reference network in periods $\{\bar{x}_d\}_{d = 1,\dots,365}$, where $\bar{x}_d=\left(x_{24 \cdot (d-1)+1},\dots,x_{24 \cdot (d-1)+ 24}\right)$.~After that, the~hierarchical clustering (as~implemented in the Python package Scikit-learn.cluster.AgglomerativeClustering using Ward's algorithm) is applied to aggregate the days in a fixed number, $k$, of groups (not necessary consecutive, i.e.,\ in a non-chronological order). In the first step, this clustering method aggregates two days with minimum deviation in their data sets $\Vert \bar{x}_d - \bar{x}_d^{\prime} \Vert$, where $\Vert \cdot \Vert$ indicates the Euclidean norm. Then~it aggregates the two clusters that minimizes the total within-cluster variance according to Ward's linkage criterion~\cite{Wa63}. This procedure is repeated until the desired number of clusters $k$ is reached.  The~coupling clustering is chosen since it performs better than other clustering methods in the context of \acsp{tsam}, cf.~\cite{Ko18b}. The~choice of daily periods has the advantage of better approximating data with daily patterns, like the load and the solar generation, but at the price of approximating less accurately the wind generation, cf.\ \cite{Pi18}. The~\emph{chronological clustering} uses the same principle as the coupling clustering with the additional condition to cluster only consecutive hours, i.e.,\ chronologically, of the time series of the reference network $\{x_t\}_{t = 1,\dots,8760}$~\cite{Pi18}. Notice that the clustering is applied to load and generators in block, since it produces a more drastic simplification of the reference network in contrast to what would happen if the cluster would be done for each generator and the load independently.

\begin{figure}[H]
\centering
\includegraphics[width=.98\linewidth]{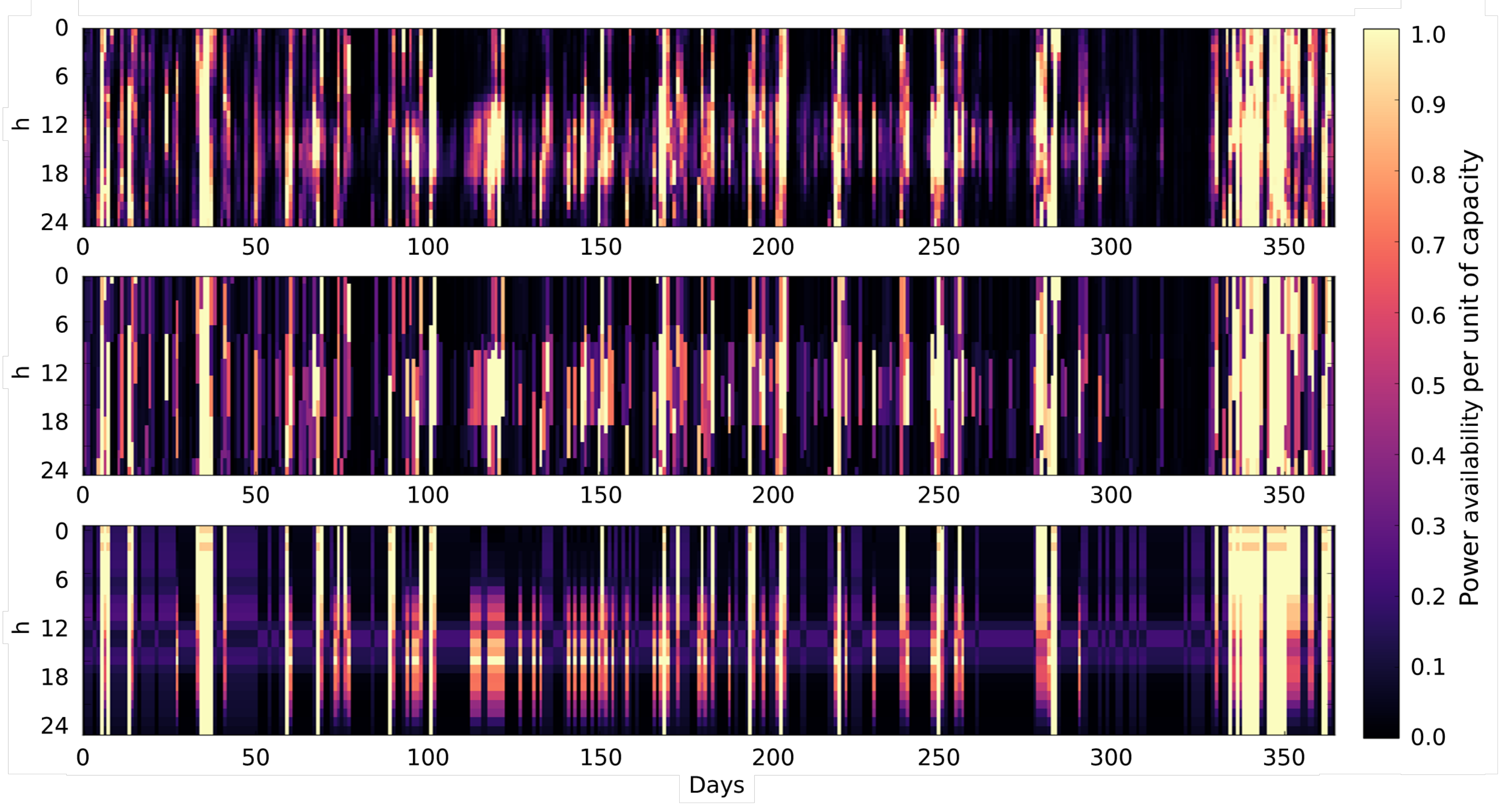}
\caption{Time series for a wind generator with corresponding power availability. The~graphics have a row for each hour of a day and a column for each day of the year. On top the reference series, in the middle the chronological clustering using 120 representative hours and at the bottom the coupling clustering using 5 representative days.}\label{fig_tsam5wind}
\end{figure}

After the clustering step, the~\emph{medoid} of each cluster is chosen as representative period, that is, the~element of the cluster that is closer to the mean value of all the cluster components. Notice~that, other clustering algorithms choose the \emph{centroid} of the cluster instead, which is the mean value of all the elements in the cluster.~However, in the context of \acsp{tsam}, the~representative periods produce better approximations than the centroid periods since they do not discard extreme periods, cf.\ \cite{Ko18b}. A~visualization of the two clustering methods is shown in Figure \ref{fig_tsam5wind} for 5 representative days (respectively 120 h) of a wind generator. Since the chronological clustering is not forced to use daily patterns (like the coupling clustering), it is able to capture more accurately the abrupt changes in the wind intensity. On the other hand, the~load and the solar energy are very poorly approximated by the chronological clustering when not enough representative hours are considered.

\subsubsection{Data Rescaling}

After the clustering, the~data needs to be rescaled or normalized (to reverse the preprocessing done in Section \ref{sec_pretreat}), cf.\ \cite{Na16} (Step~6).

\subsubsection{Adapting the Linear Problem}\label{sec_couple}

The chronological method \cite{Pi18}, reduces the equations of the linear problem. This is achieved by using just the equations corresponding to the representative hours and defining the weighting $\omega_t$ at a representative hour $t$ as the number of components represented by $t$. The~coupling method~\cite{Ga18} (Method~M1), makes the same reduction to the representative days, but also requires to keep the time linking constraints regarding the state of charge, Equation (\ref{c9}), for every single hour of the year, i.e.,~\emph{coupling} the days in the right chronological order. This is done using a map of the distribution of each representative day through the year. Notice that, the~chronological clustering reduces all the time dependent variables, equalities, and inequalities from \num{8760} to the number of clusters $k$. On the other hand, the~coupling clustering has to keep all the \num{8760} equations corresponding to the state of charge (and the variables involved).

\subsection{Indicators}\label{sec_indicators}

In order to compare the \acs{lopf} of the temporally aggregated networks to the reference network, we use the so called \acfi{aoe} \cite{Pi18}, which has the following general formula:
\begin{equation}\label{eq_aoe}
\left(\frac{\bar{z} - z}{\bar{z}}\right)\times 100 \%,
\end{equation}
where $z$ and $\bar{z}$ are, respectively, the~objective values of the network before and after the temporal aggregation. Recall that the objective value is the annual cost (which strongly depends on the built storage capacities). The~same formula applied to the computational time calculates the \acfi{atr}.

The \acs{aoe} and the \acs{atr} are the most common indicators used to evaluate the trade\nobreakdash-off between the accuracy and the complexity of an energy model, cf.\ \cite{Ga18,Ko18,Ko18b,Pi18}.~More conceptual indicators are considered in~\cite{Mer19}, but they only consider temporal aggregations that do not change the model output (Section~6.2.1,~\cite{Mer19}). In order to compare the accuracy of the clustering before the optimization, following~\cite{Na16}, we will use the \emph{Pearson correlation}. If $\{l'_{i,t}\}_{i\in I, 0\leq t\leq 8760}$ denotes the time series at node $i$ of the load aggregated time series, the~Pearson correlation is defined~as:
\begin{equation}\label{eq_Pearson}
{\mbox{\rm Corr}}_{i}= 1 - \frac{\sum_{t=1}^{T}\left(l_{i,t}-\bar{l}_{i}\right)\left(l'_{i,t}-\bar{l'}_{i}\right)}{\sqrt{\sum_{t=1}^{T}\left(l_{i,t}-\bar{l}_{i}\right)^2}\sqrt{\sum_{t=1}^{T}\left(l'_{i,t}-\bar{l'}_{i}\right)^2}},
\end{equation}
where $\bar{l}_{i}$ and $\bar{l'}_{i}$ are the respective mean values of the time series. The~Pearson correlation of the load, ${\mbox{\rm Corr}}_{\mbox{\rm{load}}}$, is the average value of all ${\mbox{\rm Corr}}_{i}$ for all $i\in I$. It is similarly defined for the time series of solar and wind energy.

Other indicators that were considered for evaluating the accuracy of the clustering were the normalized root square error, the~covered variability (Equation~(12), \cite{Na16}), the~skewness and the Kurtosis. All these indicators (also applied to the duration curves of the time series, cf.\ Appendix~B, \cite{Na16}) were tested in the preparation of this paper, but finally discarded, since they gave no further hints than the Pearson correlation.

\subsection{Solver and Hardware}\label{sec_gurobi}

We use the optimization solver Gurobi with an academic licence \cite{gurobi}, and with the Barrier solving method (with deactivated crossover).~This method was selected to enable parallel computations. We~fixed the convergence tolerances to \num{e-4}.~The computations were conducted on a server with 48~Intel(R) Xeon(R) Gold 6128 CPUs @ \SI{3.40}{\giga\hertz} and \SI{3}{\tera\byte} RAM.

\section{Test Case Optimization}\label{sec_test_opti}

\begin{figure}[H]
\centering
\includegraphics[width=0.95\linewidth]{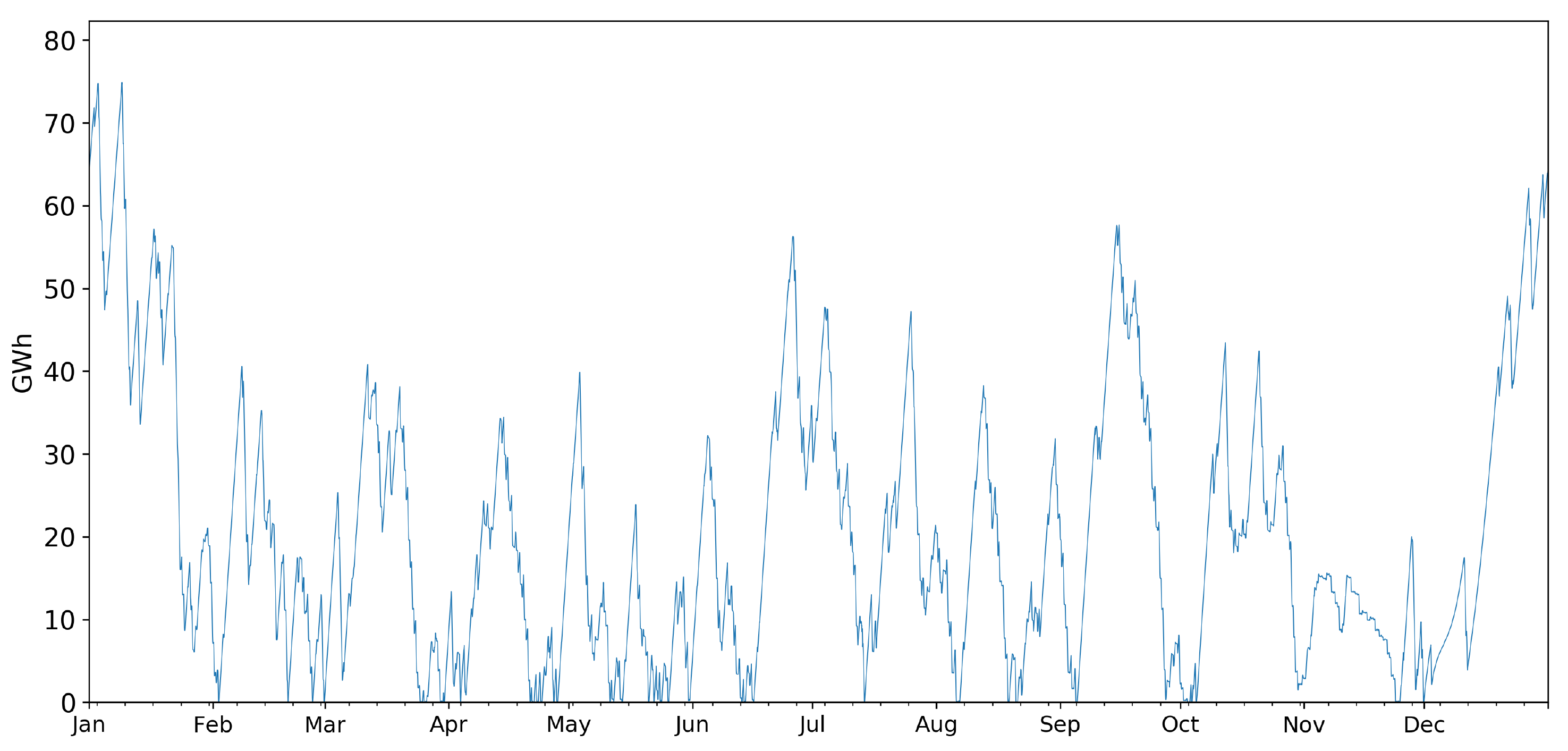}
\caption{Global state of charge for the non-aggregated test case.}\label{fig_soc}
\end{figure}

The \acs{lopf} of the test case (without temporal aggregation) results in a cost value of \SI{103}{\million\EUR} and it is calculated in ca.\ \num{110} min. All the battery storage units are optimized to a capacity around \SI{0.25}{\mega\watt} and \SI{1.5}{\mega\watthour} and most of the compressed hydrogen underground storage caverns are around \SI{8.7}{\mega\watt} and \SI{1.5}{\giga\watthour}. Notice that this capacity is similar to the capacity established by other research papers for similar scenarios, cf.\ \cite{El16} (page~53). The~total storage capacity of the non-aggregated test case sums up to \SI{74.87}{\giga\watthour}. At the beginning of the year, the~\acs{lopf} considers an almost full global state of charge (\SI{64.36}{\giga\watthour}), which is reasonable since there are no costs for the energy stored the previous year, see Figure~\ref{fig_soc}. On the other hand, the~state of charge is imposed to be the same for each storage unit at the beginning and at the end of the year.

The total share of dispatched energy carriers is shown in Table \ref{table_carriers_share}.~The energy  is highly dominated by wind energy ({71.58}\%) and the total renewable energy share amounts to {90.98}\%.~This~is a consequence of the fact that wind energy does not have management costs, that the storage potential is very large and economically competitive, and that the wind capacity in the region of Schleswig-Holstein is meant for exportation to less windy regions (but in our model we set it only for internal consumption).
\unskip

\begin{table}[H]
\caption{Share of dispatched energy by carriers (\SI{}{\tera\watthour}).}
\label{table_carriers_share}
\centering
\begin{tabular}{cccccc}
\toprule
\bf{Wind} & \bf{Solar} & \bf{Gas} & \bf{Waste} & \bf{Coal} & \bf{Biomass} \\
\midrule
\num{9.49} & \num{1.38} & \num{0.22} & \num{0.05} & \num{0.92} & \num{1.19} \\
\num{71.58}\% & \num{10.40}\% & \num{1.65}\% & \num{0.41}\% & \num{6.96}\% & \num{9.00}\% \\
\bottomrule
\end{tabular}
\end{table}

\begin{figure}[H]
\centering
\includegraphics[width=0.93\linewidth]{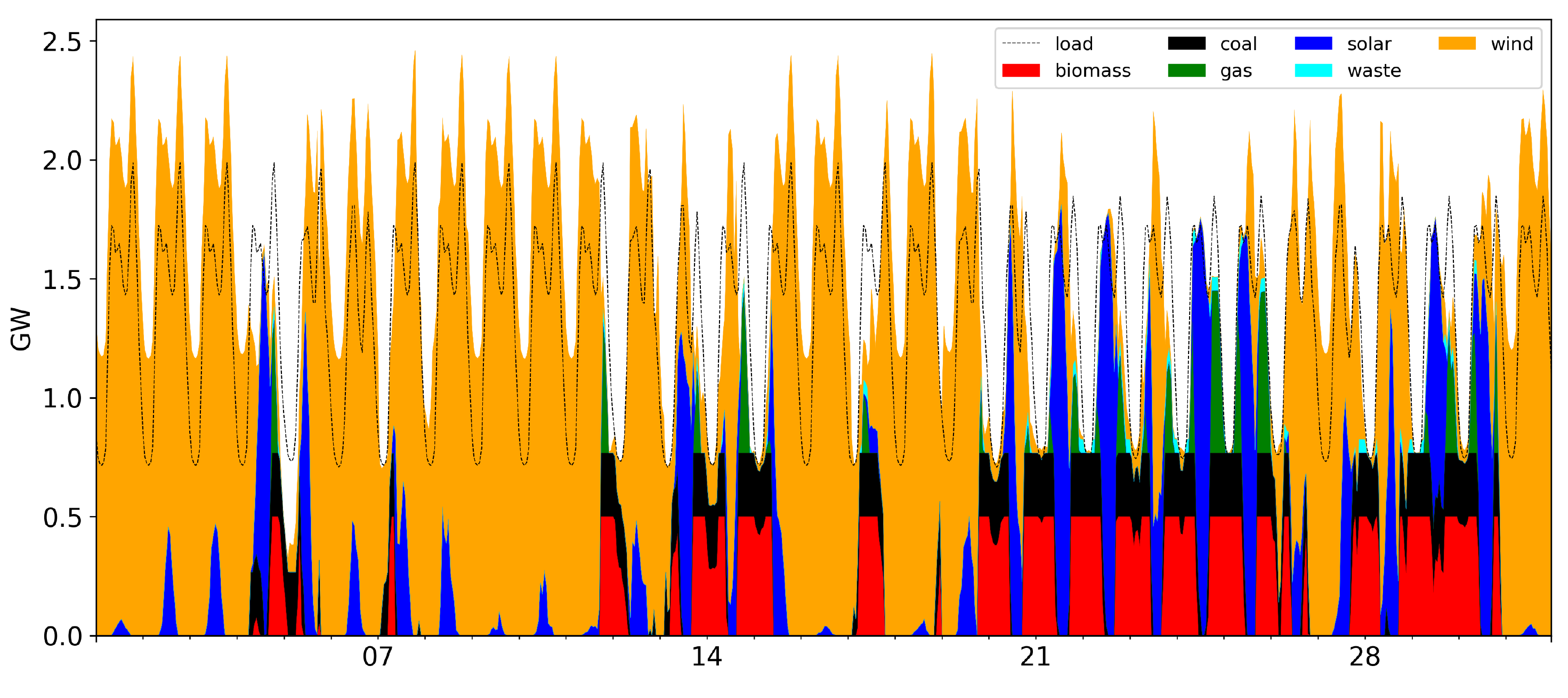}
\caption{Overall generated power dispatched in March stacked by carrier.}\label{fig_stacked_gen_and_power-load-storage3}
\end{figure}

Although the load is mostly covered by wind and solar, in less windy periods there is not enough renewable energy generation to cover the energy demand, even using all the available stored energy. \mbox{As an example}, we plot the generated power for March stacked by carrier in Figure \ref{fig_stacked_gen_and_power-load-storage3}. There is a big power curtailment of {74.11}\% of wind energy and of {44.33}\% of solar energy, which is neither used nor stored. This~is a consequence of the fact that the installed generation capacity is fixed, but the storage is optimized in order to minimize the costs. It is also magnified by large wind generation capacity of the region. So~adding new transmission lines to adjacent regions or increasing the capacity of the existing lines would certainly decrease the curtailment.

\section{Time Series Aggregation Efficiency}\label{sec_tsam_efficiency}

\begin{figure}[H]
\centering
\includegraphics[width=0.98\linewidth]{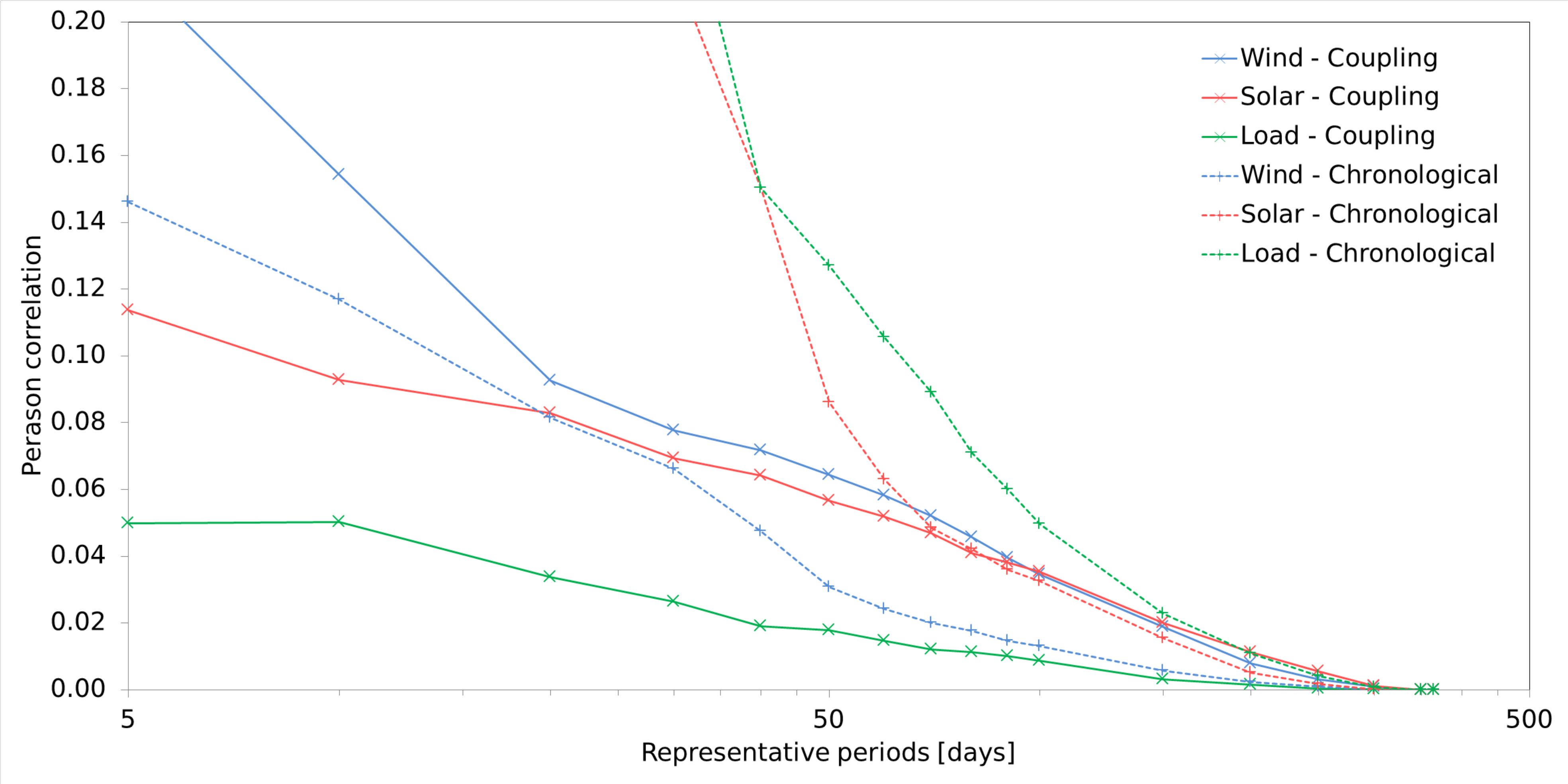}
\caption{Pearson correlation for the coupling and chronological clustering for the load, wind and solar energy, using different representative periods.}\label{fig_Pearson}
\end{figure}

The Pearson correlation of the aggregated time series is shown in Figure \ref{fig_Pearson}. This graphic shows how the different time series get better approximated with more representative periods. Note that the coupling clustering approximates the load and the solar energy better than it does for wind energy (since the latter does not show a daily pattern). The~chronological clustering, on the other hand, approximates the wind energy better and it needs relatively a lot of representative hours to be able to approximate the solar energy and the load. Considered altogether, both clustering methods monotonically improve the approximation with the increase of representative periods, showing a good potential for the application that this article~pursues.

\section{Time Reduction and Error}\label{sec_time_red_lopf}

\begin{figure}[H]
\centering
\includegraphics[width=0.98\linewidth]{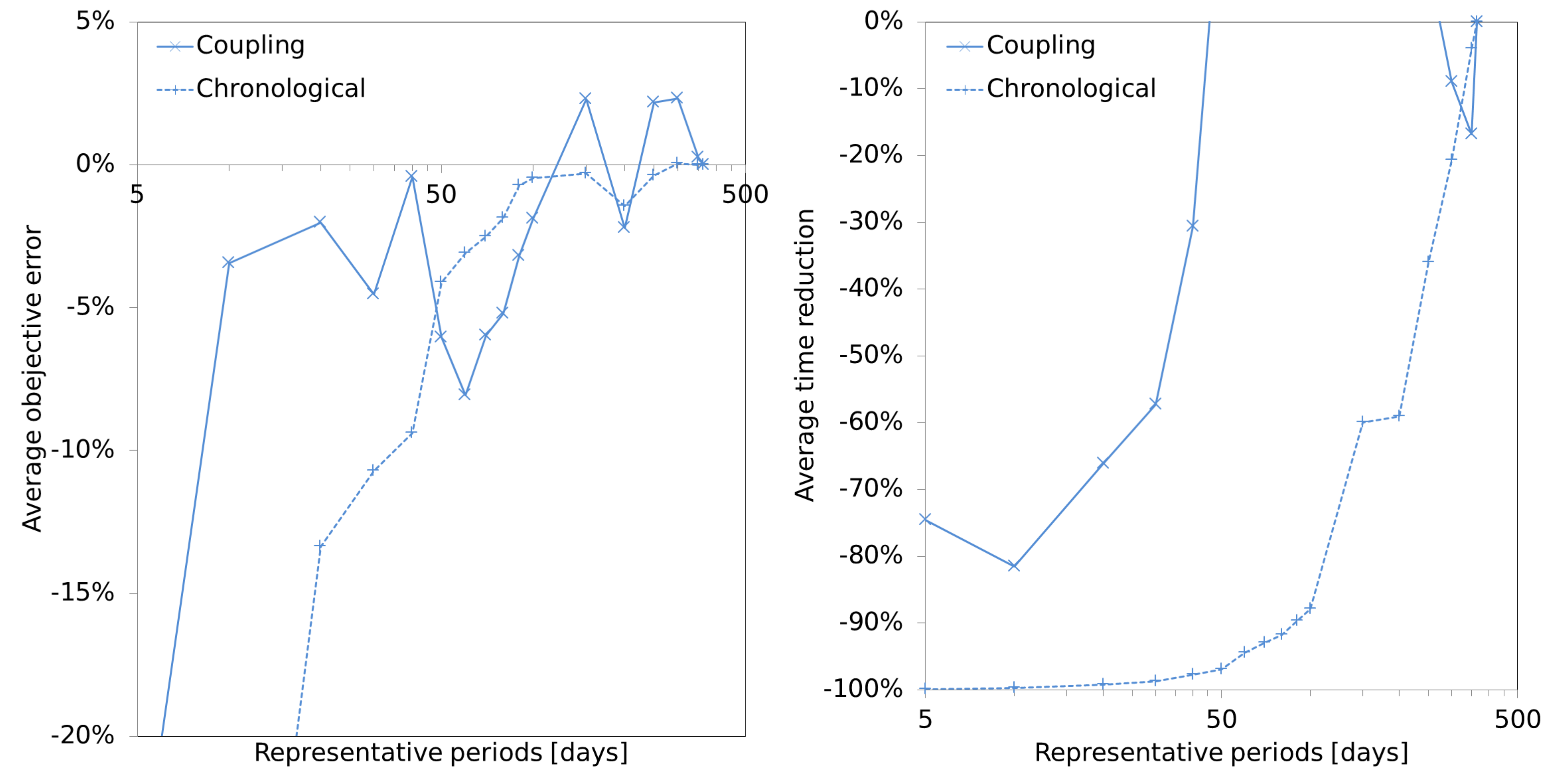}
\caption{Average objective error (\textbf{left}) and average time reduction (\textbf{right}) for the coupling and chronological clustering using different representative periods.}\label{fig_error_time}
\end{figure}

\begin{figure}[H]
\centering
\includegraphics[width=0.98\linewidth]{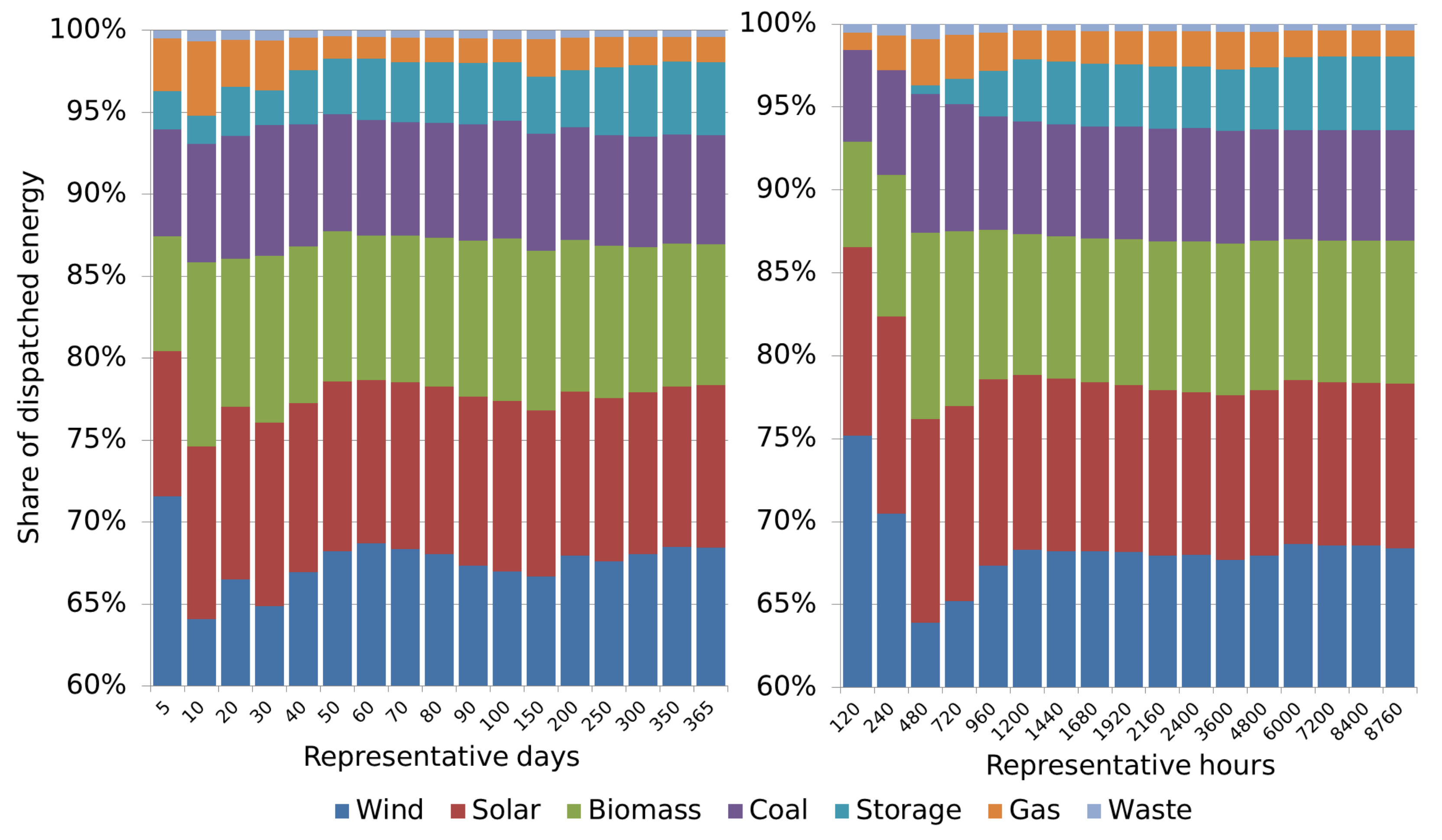}
\caption{Share of dispatched energy for the coupling method (\textbf{left}) and the chronological method (\textbf{right}). Notice that wind energy is always above the {60}\% share.}\label{fig_shareofcarriers}
\end{figure}

\begin{figure}[H]
\centering
\includegraphics[width=0.85\linewidth]{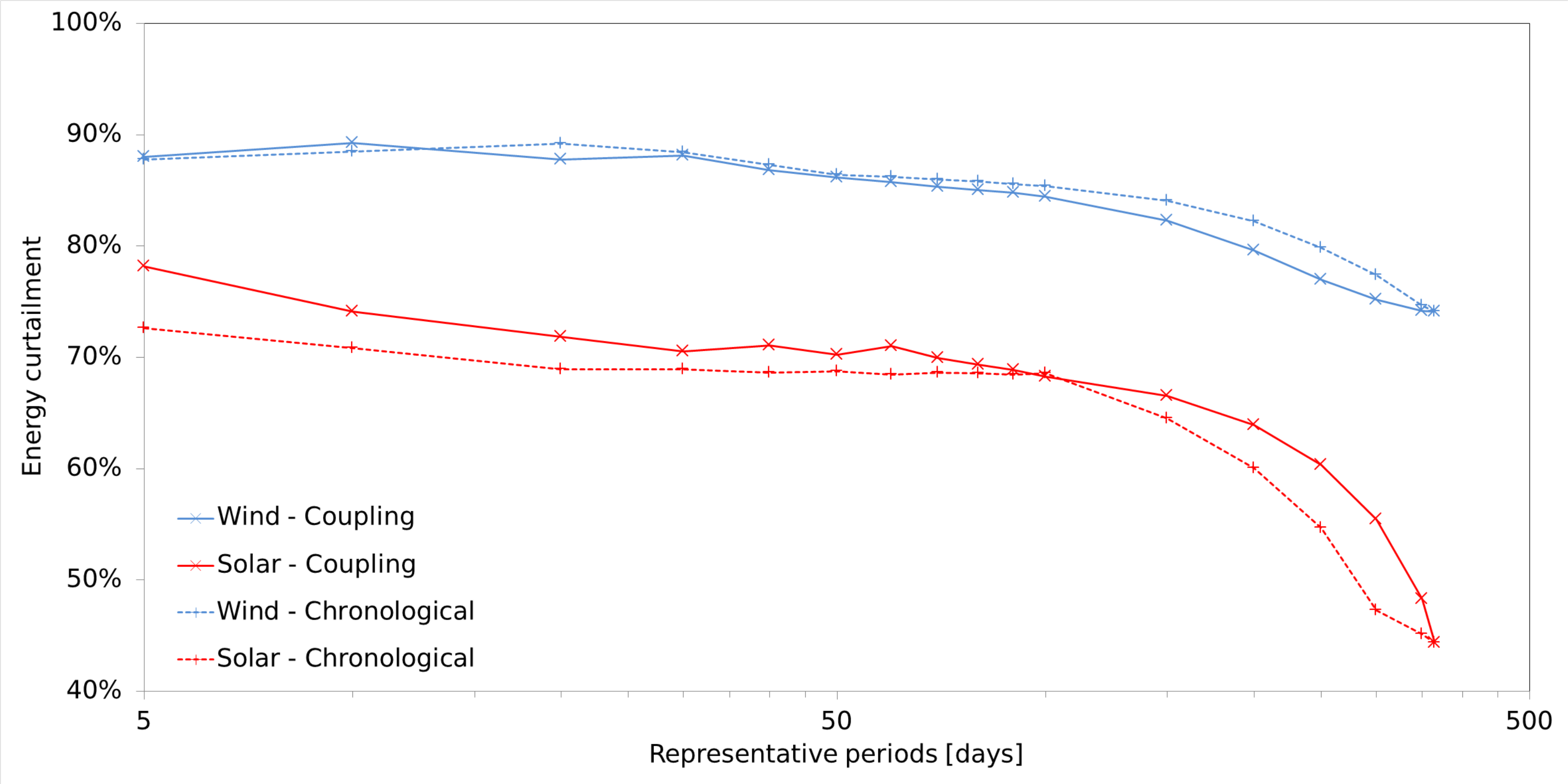}
\caption{Curtailment of wind and solar energy for the coupling and chronological clustering using different representative periods.}\label{fig_curtailment}
\end{figure}

The \ac{aoe} and the \ac{atr} for an increasing number of representative periods for the different \acsp{tsam} are shown in Figure \ref{fig_error_time}.~With the chronological clustering, the~error is reduced almost monotonically to zero, although it is very high when we consider just few representative hours.~Most notably, the~computational time is reduced very significantly. For~instance, with 2400 representative hours we have an \acs{aoe} of less than 1\% and a time reduction of 88\%. Notice that, the~\acs{aoe} is mostly negative, meaning that the global costs of the system, i.e.,~the~objective value, is underestimated. This means that the storage capacity needed by the system is underestimated, see Figure~\ref{fig_shareofcarriers}. This is probably due to the inaccurate representation of the variability of the available wind energy, which is also the most curtailed, see Figure \ref{fig_curtailment}.

\unskip
\unskip

On the other hand, the~coupling clustering shows a more chaotic error convergence and the time reduction is not considerable. In fact, above 50 representative days it takes more time to compute than the reference problem, which makes this temporal aggregation useless. This might seem a contradiction with the fact that the temporal aggregation creates a smaller linear problem, but we must also account for the modifications of the equations representing the state of charge, Section \ref{sec_couple}, which increase the relations between the variables and, thus, the~complexity of the problem. Even if the \acs{aoe} is very small for 40 representative days, it only reduces the computational time by {30}\%. Even worse is the fact that the error does not reduce gradually nor monotonically, hence making it difficult to decide for a convenient number of representative days to use.

Different methods to decide the number of representative periods (before the temporal clustering) have been proposed, cf.\ \cite{Na16} (Section~5.1) or \cite{Fa14a}. However this is a challenging issue and there are no conclusive results. Even more, the~methods available do not help in our case, since there is no correlation between the error of the clustering (Figure \ref{fig_Pearson}) and the \acs{aoe} (Figure \ref{fig_error_time}). The~only reliable process is to progressively compare the objective value of the network aggregated to $k$ nodes with the network aggregated to $k-1$ nodes and stop when the difference between the objective values is small enough. Nevertheless, it can only be helpful with the chronological clustering approach. In the case of the coupling clustering, because of its chaotic behaviour, this method does not give a good indicator of the error convergence, see Figure \ref{fig_error_time}.

The chaotic behaviour of the \acs{aoe} obtained for the coupling clustering was not mentioned in the article where the method was introduced \cite{Ga18}. This may be due to the much smaller network used (see~Section \ref{sec_size}) and that the optimization is carried out using a \acs{milp}. This behaviour can be, nevertheless, observed in \cite{Ko18} (Figure~6) for an island system with two storage units. However, contrary to our results, in this reference, the~computational time is always reduced by the coupling clustering. As we will see in Section \ref{sec_size}, this is a consequence of the fact that they use a much smaller network compared to ours. In \cite{Ko18b}, they also introduce two alternative implementations of the coupling clustering aimed at further reducing the computational time. However, when tested with our test case, we obtain a similar \acs{atr} (Figure~\ref{fig_error_time}) as for the original method formulation in \cite{Ga18}.

The results of the chronological clustering are consistent with the results in \cite{Pi18} (Figure~3), where the method was introduced.~However, in that article, it is compared to the clustering of (non\nobreakdash-adjacent) representative days without changing the time linking constraints tracking the state of charge (Equation~(\ref{c9})). That makes the error much bigger than the coupling method presented here (to~the point of making the clustering methods useless as already observed in \cite{Ko18b,Ga18}). The~advantage of the methods compared in \cite{Pi18} is that the computational time reduction is similar in all of them, cf.~\cite{Pi18}~(Figure~4), and no anomaly (as in Figure \ref{fig_error_time}) is observed. What our results show is that the coupling clustering can actually give very good approximations, but in an unpredictable way and not necessarily reducing the computational time.

\section{Influence of the Power Network Size}\label{sec_size}

The results of Section \ref{sec_time_red_lopf} for our test case were also confirmed by repeating the calculations for other power networks of similar or bigger size constructed from the \acs{openego} network, {cf.}\ \cite{openego}, and the PyPSA-Eur network, {cf.} \cite{pypsaEur}. However, if the size of the power network is reduced, the~computational time reduction of the coupling clustering has a better behavior. In order to reduce the network we use the spatial clustering method implemented in \cite{pypsa}.~Roughly, it first uses a $k$-mean clustering algorithm to cluster the buses into $k$ groups and chooses the centroid as the representative of each cluster. Then~the lines connecting the same clusters have their capacity aggregated into a single line connecting the respective centroids and all the generators, loads and storage units attached to the buses of one cluster are attached to the centroid.

\begin{figure}[H]
\centering
\includegraphics[width=0.8\linewidth]{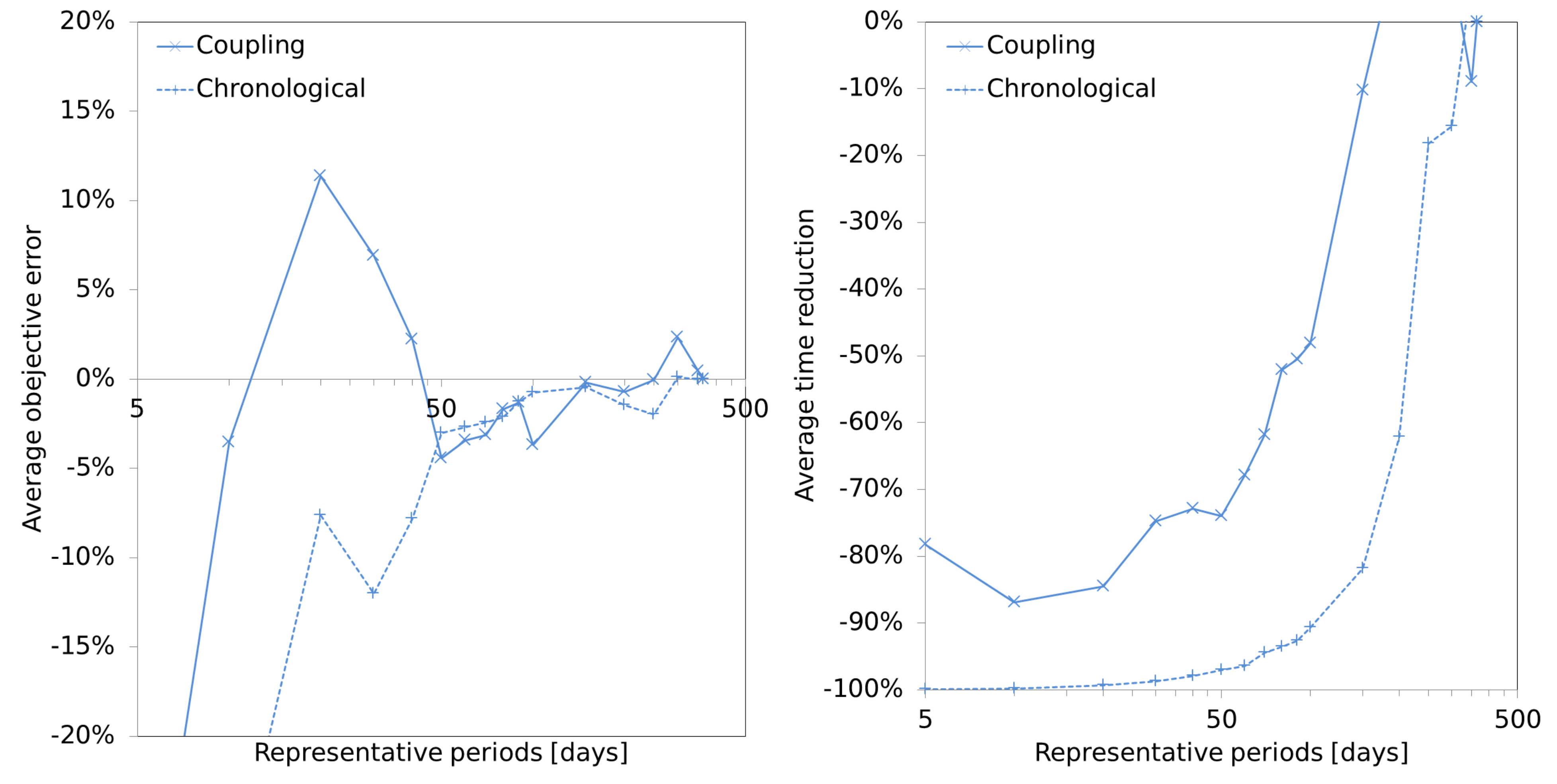}
\caption{Average objective error (\textbf{left}) and average time reduction (\textbf{right}) for the 10 nodes test case for the coupling and chronological clustering using different representative periods.}\label{fig_error_time10}
\end{figure}

In Figure \ref{fig_error_time10}, we show the \acs{aoe} and the \acs{atr}, respectively, for the test case of a power network spatially clustered to 10 nodes. As in the original test case, the~\acs{aoe} of the coupling method still has a rather chaotic behavior, whereas the chronological clustering behaves in a better way. However, in~this case, the~coupling clustering does reduce significantly the computational time, although not as much as the chronological clustering. For instance, using just 50 representative days (correspondingly~1200~h) both algorithms have a similar \acs{aoe}, but the \acs{atr} of the chronological clustering is of {96}\% whereas for the coupling clustering it is of {74}\%.

It is also interesting to stress the remarkable computational simplification of the spatial clustering. The~computational time is reduced by more than {95}\% and the error is just of {0.01}\%. Hence, if we are not interested in the spatial distribution of the storage planning, the~spatial clustering is a much more efficient method to reduce the computational time than the \acsp{tsam}, at least in the system under study with a high dependency on wind energy generation, cf.\ \cite{Pri19} (page~21).

\section{Influence of Parallel Computation}\label{sec_CPU}

The Barrier algorithm from the Gurobi solver with the academic license allows for running parallel computations with up to 8 CPU threads. To evaluate the benefits of using parallel computation combined with the \acsp{tsam}, we repeat the previous computations from Section \ref{sec_time_red_lopf}, previously done with 4 CPU threads, with 1, 2 and 8 CPU threads. However, it is also important to notice that the computational time reduction is known not to increase linearly nor indefinitely when increasing the number of threads, cf.\ \cite{ParallelEncyl} (page~55).

\begin{figure}[H]
\centering
\includegraphics[width=.95\linewidth]{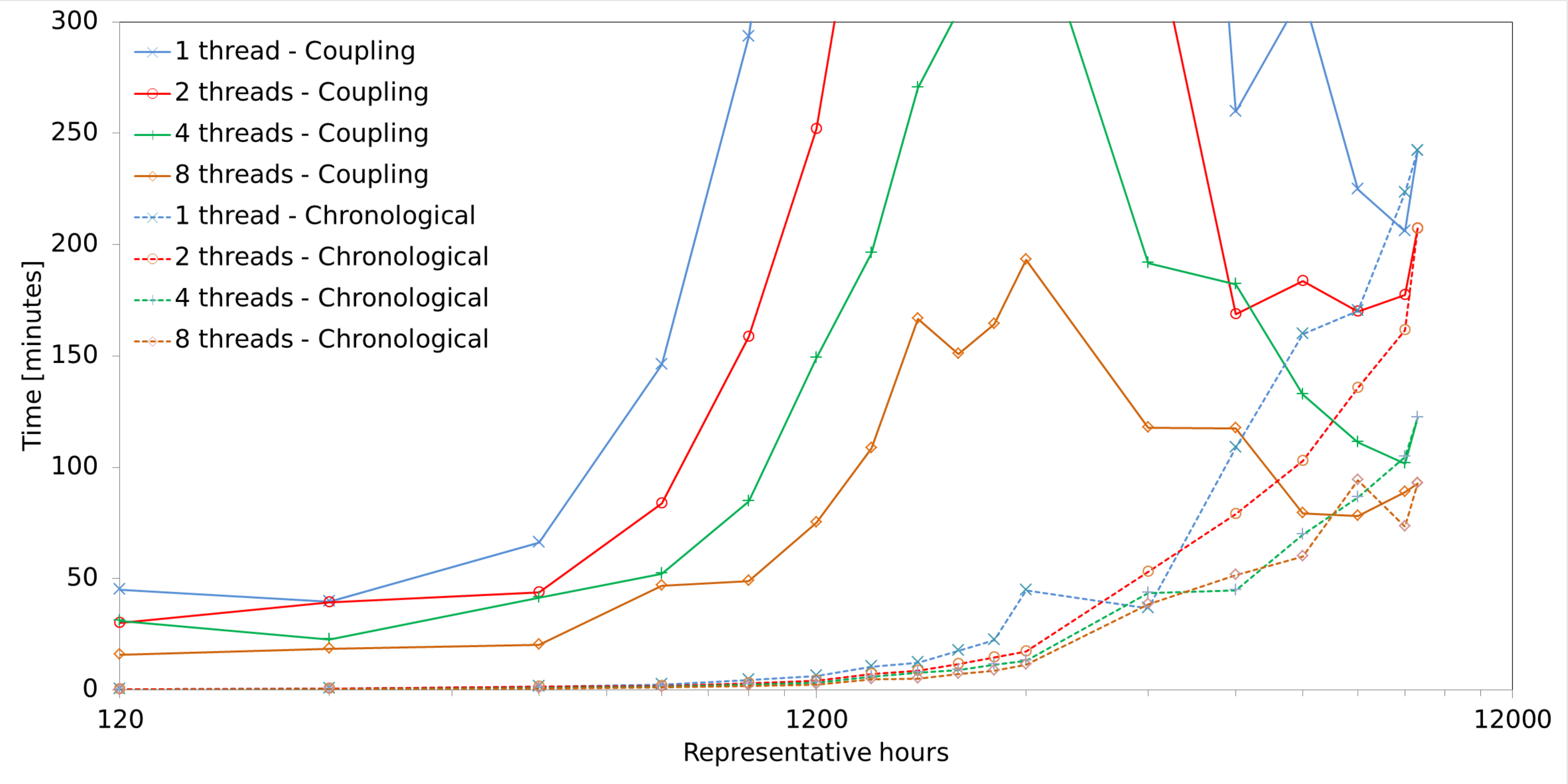}
\caption{Computational time for the coupling and chronological method using 1, 2, 4, and~8 CPU~threads.}\label{fig_CPU_parallel}
\end{figure}

The absolute computational time for the coupling clustering and the chronological clustering as well as the different number of CPU threads used are shown in Figure \ref{fig_CPU_parallel}. For the coupling clustering, there is no time reduction for 50 representative days or more (like for the case of 4 CPU threads in Figure \ref{fig_error_time}). For 40 days, the~time reduction achieved by the parallel computation (comparing 1 thread to 8 threads) is of a similar magnitude of that of the clustering with 4 threads. On the other hand, for~the chronological clustering, the~advantages of parallel computation are less remarkable, since the time reduction achieved by the temporal clustering in Figure \ref{fig_error_time} is bigger than the one achieved by the parallel computing.

\section{Conclusions}\label{sec_conclusion}

This article evaluates two time series aggregation methods to reduce the computational time of the linear optimal power flow computation for a power grid highly depending on wind generation and with storage planning: the coupling clustering of independent days and the chronological clustering of consecutive hours. The~fact that the storage planning include both intra-day and inter-day storage makes other existing methods (with no coherent state of charge equations) inappropriate. The~use of the hierarchical clustering instead of other clustering methods is based in a consistent better behavior in the consulted literature.

\newpage
The present paper can help to choose a complexity reduction technique, with a remarkable impact in the time reduction of the optimization process. This would allow the optimization of larger storage planning energy models, which can help in making important decisions in energy policy to facilitate the transition into a carbon\nobreakdash-neutral energy system.

The chronological clustering gave the best results in terms of computational time reduction. Also~the error obtained converges rather monotonically to zero when increasing the number of typical hours, hence having a more predictable behavior.

The coupling day clustering is not always able to reduce the computational time and the error does not converge smoothly to zero, hence making it difficult to use in practice. This behavior is expected with power networks of a similar (or bigger) size and characteristics to the test case. However, for much smaller networks, the~coupling method improves considerably. The~reason behind the bad performance of the coupling method is that the reduction of equations was canceled out by the intertwining of the time linking constraints keeping track of the state of charge of the storage units. Furthermore, the~use of more CPU threads does not contribute significantly to the computational time reduction already achieved by the \acsp{tsam}, specially with the chronological method.

It is also worth mentioning, that the spatial clustering is a much more efficient method to reduce the computational time than the \acsp{tsam}, which agrees with other sources, like \cite{Pri19} (page~21). That is in case that the spatial distribution of the storage planning is not relevant for the research study.

The error of the clustering method itself (Figure \ref{fig_Pearson}) is not sufficient to assess the error of the objective value (Figure \ref{fig_error_time}), as observed also in other studies, cf.\ \cite{Cao18} (p. 20) or \cite{Tei19} (p. 11). This makes it difficult to anticipate the best number of representative periods.
Hence, there is no optimal temporal clustering method for all possible and addressed research questions. But for the special case of an energy system highly depending on wind energy or in general time series with rather non-regular pattern, we suggest the use of the chronological clustering. In this case, the~number of representative periods can be identified by using the convergence of the optimal objective value as we increase the number of representative periods.

The results of this article can benefit model energy projects to choose a complex reduction technique for the optimization of an energy model. This is a crucial step in the optimization process, since some techniques would result in unacceptable errors or, as we highlighted here, in no computational time reduction at all. To put it concisely, we showed that the chronological clustering performs the best, when dealing with large storage planning energy network models with time linking constraints. This clarifies some discussions in previous publications by comparing the two most valued methods that keep track of the storage units state of charge. We also showed that, if the spatial distribution of storage is not necessary, the~temporal clustering can be combined very efficiently with a spatial clustering.


\vspace{6pt}



\authorcontributions{{Conceptualization}, O.R.; Investigation, O.R.; Methodology, O.R. and J.B.; Writing---original draft, O.R.; Writing---review \& editing, O.R. and J.B. All authors have read and agree to the published version of the~manuscript.}

\funding{This research was funded by the open\_eGo project, which was supported by the
German Federal Ministry for Economic Affairs and Energy under the funding code: BMWi 0325881D.}

\acknowledgments{The authors would like to thank {Wided Medjroubi} for suggesting the research topic and giving continuously feedback through the research process. The~authors also want to thank Jonas H\"{o}rsch, Leander Kotzur and Bruno Schyska for fruitful conversations about linear optimization of power systems. The~authors acknowledge Gurobi Optimization Inc. for providing an academic licence.}

\conflictsofinterest{The authors declare no conflict of interest.}

\newpage
\abbreviations{The following abbreviations are used in this manuscript:\\ \null

\noindent
\begin{tabular}{@{}ll}
\acs{aoe} & \acl{aoe} \\
\acs{atr} & \acl{atr} \\
\acs{dlrve} & \acl{dlrve} \\
\acs{etrago} & \acl{etrago} \\
\acs{lopf} & \acl{lopf} \\
\acs{milp} & \acl{milp} \\
\acs{nep} & \acl{nep} \\
\acs{openego} & \acl{openego} \\
\acs{oep} & \acl{oep} \\
\acs{pypsa} & \acl{pypsa} \\
\acs{tsam} & \acl{tsam} \\
\end{tabular}
}

\reftitle{References}

\end{document}